\documentclass{MITcsail}

\title{From OECD to India: Exploring cross-cultural differences in perceived trust, responsibility and reliance of AI and human experts}

\author
{Vishakha Agrawal~$^{1}$\footnote{Correspondence E-mail: vishakha.agrawal09@gmail.com \\
Permission to make digital or hard copies of part or all of this work for personal or classroom use is granted without fee provided that copies are not
made or distributed for profit or commercial advantage and that copies bear this notice and the full citation on the first page. Copyrights for third-party
components of this work must be honored. For all other uses, contact the author(s). \\
© 2023 Copyright held by the author(s).},  Serhiy Kandul~$^{2}$, Markus Kneer~$^{2}$, Markus Christen~$^{2}$\\
\vspace{1em} 
\normalfont{\small $^{1}$Independent Researcher, India}\\
\normalfont{\small $^{2}$University of Zurich, Switzerland}\\
}

\begin{document}

\maketitle

\begin{abstract}
AI is getting more involved in tasks formerly exclusively assigned to humans. Most of research on perceptions and social acceptability of AI in these areas is mainly restricted to the Western world. In this study, we compare trust, perceived responsibility, and reliance of AI and human experts across OECD and Indian sample. We find that OECD participants consider humans to be less capable but more morally trustworthy and more responsible than AI. In contrast, Indian participants trust humans more than AI but assign equal responsibility for both types of experts. We discuss implications of the observed differences for algorithmic ethics and human-computer interaction.
\end{abstract}

\section{Introduction}

Artificial intelligence (AI) is becoming increasingly important in various applications from HR decisions \citep{park2021human} and processing of clinical data \citep{yang2019unremarkable,lee2021a} to writing essays and co-creation of art \citep{li2020empowering}.
In fact, the technology has advanced to the stage that AI successfully performs tasks believed to be better handled by humans. A recent news article \citep{Else:2023} shows that scientists cannot distinguish essays written by ChatGPT from human-written texts. 

Although AI-based decision support system often demonstrate superior results, acceptability  of AI in certain contexts remain low \citep{Burton2020, jussupow2020we}. However, in other contexts, people seem to value AI-generated advice more than a human advice \citep{logg2019algorithm}. The literature has therefore looked for factors behinds people's preferences for and against AI (and human) decision support. Examples include the nature of the task (e.g., degree of task subjectivity \citep{castelo2019task}), need for (human) discretion and differentiated treatment \citep{Bai:2021, Uhl:2022}, explainability and transparency of AI \citep{Ferrario:2022, Leichtmann:2023}, or framing of the expert power and autonomy \citep{Hou:2021}.

Another line of research links the variation in attitudes towards AI to \textit{cultural differences} \citep{lee2021who, kim:2022}. \cite{lee2021who}, for example, argues that minority social groups might have developed mistrust in human (public health) systems due to historical discrimination and therefore are not averse to algorithms. \cite{Haring:2014} proposes that different level of exposure and public image of the technology drive cultural differences in attitudes towards robots in Europe and Japan. \cite{Tubadji:2021} compares adoption rates of AI-based banking advice systems and suggests that cultural differences in attitudes towards risk and uncertainty is the main driver of cross-country differences in adoption rates.
Although a few studies highlighted potential cultural difference in specific contexts, overall the research on AI perceptions predominantly relies on Western samples \citep{Vanberkel:2023}. This calls for more cross-cultural analysis.

The workhorse of much cross-cultural research contrasting Western and Eastern countries is \cite{Hofstede:80}'s cultural dimensions (for a recent review of studies implementing the framework, see \cite{Beugelsdijk:2017}); some call for extensions, see e.g. \cite{Nakata:2009}, for social robotics in particular, see \cite{Ornelas:2022}. Research shows that cultural differences across the individualism-collectivism dimension, which defines the relative importance of individual and community values and goals, drives the differences in the perception of, and interaction with, others \citep{Brewer:2007}. This difference is also considered one of the major factors for the explanation of West-East differences in human-robot interaction (for a review see e.g. \cite{Lim:2021}).

As \cite{Lim:2021} highlights, however, studies exploring West/East differences have predominantly focused on two countries, the US and Japan (see \citep[p.~1327]{Lim:2021}), with very few studies targeting Asian countries beyond Japan. Furthermore, those that do, the authors argue, suffer from a dire lack of power (see \cite[p.~1318]{Lim:2021}), India being among the countries where sample sizes have been particularly low (and thus makes it impossible to tell whether the absence of a significant difference means anything). In the paper by \cite{Lim:2021}, reviewing fifty studies from the last two decades, not a single one includes participants from India . This is a remarkable shortcoming of the literature, given that India is the second most populous country in the world, which will overtake China as the most populous country this year \citep{Ungers:2022}.

One of the few studies contrasting an Indian sample with Western participants (in this case from the US and Germany) is by \cite{Homburg:2019}. It explores attitudes towards humanoid robots, focusing on empathy and trust, as well as perceived reliability and expertise. \cite{Homburg:2019} showed participants an image of Pepper, a robot by the Softbank Corporation, whose use in service contexts is widespread. Mean ratings for all four dependent variables among Indian subjects significantly exceeded those of US and German participants \cite[p.~4777]{Homburg:2019}. In a related financial context, \cite{ramesh:22} compare AI perceptions for Global South countries and highlights the mediating role of the perceived agency of the users in evaluation of AI accountability. In the real of fair AI research, \cite{sambasivan:21} argues some critical premises of West-centric account of fairness, such as unbiased group representation in the data sets, is heavily contested in such stratified societies as India.

We argue that insights emerging from general attitude-scales are rather limited if not combined with a task-based assessment. Furthermore, it is not clear whether the scales employed, principally developed in the 1980s and focusing largely on the service context and buyer-seller relationships, are the ones with the most external validity in human-robot interaction (HRI) today. To make a serious step forward in this regard, we devised an interactive, task-based experimental paradigm complemented by a series of state-of-the art scales. Our study is one of the very first in cross-cultural HRI to test an extensive sample of Indian participants, and certainly one to use one of the most advanced designs in HRI more generally.

In this paper we consider high-stake decisions with AI or human support. We compare a OECD and a Indian population for perceived trust, responsibility attributions, and reliance on AI and human experts in this context. We are aware of the fact that this comparison concerns rather broad cultural categories. However, the OECD countries consist of highly developed countries mainly with a Western cultural background. In our sample, 73\% of the answers emerged from European countries; 23\% from Northern America and 4\% mainly from Australia, Chile and Israel so we can consider this as a good proxy for a "Western Culture". We also remind that India represents a large population with a mix of diversities in customs, rituals, traditions, language, etc., that varies from region to region within the country. 
However, on aggregate, India is known to score higher on collectivism, tolerance to power distance and restraint than Western countries.\footnote{For comparison of individual countries on specific cultural dimensions, see, e.g. https://www.hofstede-insights.com/fi/product/compare-countries/}

In our experiment, we explore three key variables of recent research in human-robot interaction. One regards \textit{trust} in an AI-driven advisor system, which picks up on a rich literature both in human-robot interaction broadly conceived (see \citep{hancock2011,lewis2018, khavas2020}) and trust in artificial intelligence \citep{glikson2020, ryan2020, siau2018}. Our approach goes beyond most extant research as it distinguishes between \textit{trust} as familiar largely from moral philosophy on the one hand, i.e. ``the attitude that an agent will help achieve an individual’s goals in a situation characterized by uncertainty and vulnerability''\citep[p.~51]{lee2004trust}, and \textit{trust in the capacities} of the AI-based application, i.e. ``...trusting that the agent is capable of completing a task'' \citep{malle2021multidimensional}. Philosophers tend to distinguish trust from \textit{reliance}, which has a more limited set of requirements – trust is usually characterized as reliance \textit{plus X}, where the further factor X can be cashed out in different terms (see e.g. \citep{pettit1995,hawley2014}). We thus tested reliance as a separate variable beyond, and distinct from, trust. In this regard, too, our study aspires to break new ground since reliance is not tested by eliciting judgment, but, rather, by a behavioral measure capturing whether people do in fact rely on the recommendations of a human or AI-driven advisor system \citep{dietvorst2015algorithm, engel:19}. Finally, we measure the extent to which people assume moral responsibility for their actions and the consequences they engender. This choice of dependent variable follows prominent calls for a novel subdiscipline of \textit{Moral Human-Robot Interaction}, which places emphasis on the moral evaluation of machine and human behavior in the context of novel technologies (see \cite{malle2015}). In fact, a number of studies have found that people are willing to deem robot actions wrong, ascribe blame to them, and attribute \textit{moral responsibility} (see e.g. \citep{malle2019, liu2022, hong2020, kneer2021playing, kneer2021can}). Part of the reason for this might reside in the fact that people are rather willing to attribute inculpating mental states, such as intention and recklessness, to AI-driven artificial agents (see \cite{stuart2021} for a review), though this is still an emerging field of research.

To make a cross-cultural comparisons of trust, responsibility and reliance, we had participants in OECD countries and India make ethical decisions upon receiving a human and AI recommendation. The task either involved minimizing casualties (defence domain) or maximizing lives saved (search and rescue domain) under time pressure and uncertainty. The design closely follows the one used in our earlier paper \citep{Tolmeijer:2022} and gives room for cultural differences: 

\begin{itemize}
    \item RQ1: How does reported trust in a human and AI expert compare across OECD and Indian populations?

    \item RQ2: Does responsibility attribution differ across OECD and Indian populations?

    \item RQ3: Are there any cross-cultural differences in reliance on human and AI experts? 

\end{itemize}

Our results indicate that OECD participants perceive AI to be more capable than humans for the given tasks, but place somewhat higher moral trust in humans. Indians trust humans more than AI but assign higher responsibility to AI agents than OECD participants.

The remainder of the paper is organized as follows: Section \ref{sec:method} present the experimental design; Section \ref{sec:results} identifies key findings; Section \ref{sec:discussion} discusses the implications and concludes.

\section{Method}
\label{sec:method}

\subsection{Experimental design}

We adopted a methodology similar to \cite{Tolmeijer:2022}. We highlight the key points from our experiment in this section. The study consisted of an experiment on the crowdsourcing platforms Prolific\footnote{\url{prolific.co}} for OECD participants (n=351) and Mechanical Turk\footnote{\url{mturk.com}} for Indian participants (n=302) to allow for a cross-cultural comparison. 

After participants accepted the task on these crowdsourcing platforms, they were sent to play a simulation on a webapp. We adapted the simulations to the respective cultural context (e.g., by changing the appearance of the avatars, using Indian or Western names and adapting the education questions to the respective naming conventions; see also Fig. 1). The narrative of the simulation was that participants were trained as operators of a drone system that had to perform either rescue or terror defence operations. They first received training and then were sent to perform four missions. There were four scenarios and each participant was randomly assigned one of these scenarios: maximise lives saved (rescue scenario) or minimise lives lost (terror defense scenario), and each of these scenarios had human-in-the-loop (the operator had to make the decision but was advised by either a AI or human expert) or human-on-the-loop conditions (the operator had to evaluate the decision made by either a AI or human expert and intervene if necessary).

The experiment consisted of the following steps (for details, we refer to \cite{Tolmeijer:2022}): First, the participants were asked to fill a consent form, enter their crowdsource platform worker ID, and pass an attention test. Then, they went through a training phase to make sure that they understand their role and task. Collecting demographic information (age, gender, education, and whether English is their native language) was integrated in the training narrative. In the course of the training phase, they received two comprehension questions, to make sure they understood the simulation environment and the tasks; failure in understanding the simulation mechanics led to exclusion. The training phase was concluded by measuring risk preference \citep{meertens2008measuring}, cognitive thinking skills, and statistical thinking skills -- latter two served as controls as the missions involved cognitive and statistical skills. After successful completion of the training, the participant were confronted with two missions with four decision problems each, once advised by a human expert and once by an AI in a random order. For every decision problem, we presented three available options. The participant had 30 seconds to decide.

The three options differed in terms of the expected number of people saved (or not killed) and respective probabilities/riskiness.
The choices presented a conflict between maximizing the expected value or maximizing the probability of helping at least somebody (or minimizing the probability of hurting somebody).

The absent for objectively clear-cut optimal choices created an interesting environment to study the effects of decision-support systems. The recommendations (human-in-the-loop) respectively decisions (human-on-the-loop) of the experts were balanced with respect to those two types of choices.

Reliance was measured by analysing whether the participants followed the expert's advice or chose a different option. At the end of each mission, the participants answered questions about how much they thought they, the AI, the human expert, the programmer of AI, and the seller of AI were responsible for the outcome on a seven-point Likert scale. Each expert provided advice with maximum probability half the time and maximum utility in the other half. The experts never gave advice with low success probability and/or low utility. After the missions, we presented the participants with two engagement questions. We also measured their affinity with technology interaction \citep{franke2019personal}, utilitarian preference \citep{kahane2018beyond}, and the trust in the AI and the human expert using the Multi-Dimensional Measure of Trust scale (MDMT) \citep{malle2021multidimensional}.
We used a 16-item version of self-reported MDMT scale comprised of the subscale for \textit{capacity trust} which captures reliability and capability of the system and the subscale for \textit{moral trust} which captures sincerity and ethicality of the system.

The participants were then sent back to the crowdsourcing platforms for payment.

\begin{figure}
\centering
\begin{subfigure}{.5\textwidth}
  \centering
  \includegraphics[width=.9\linewidth]{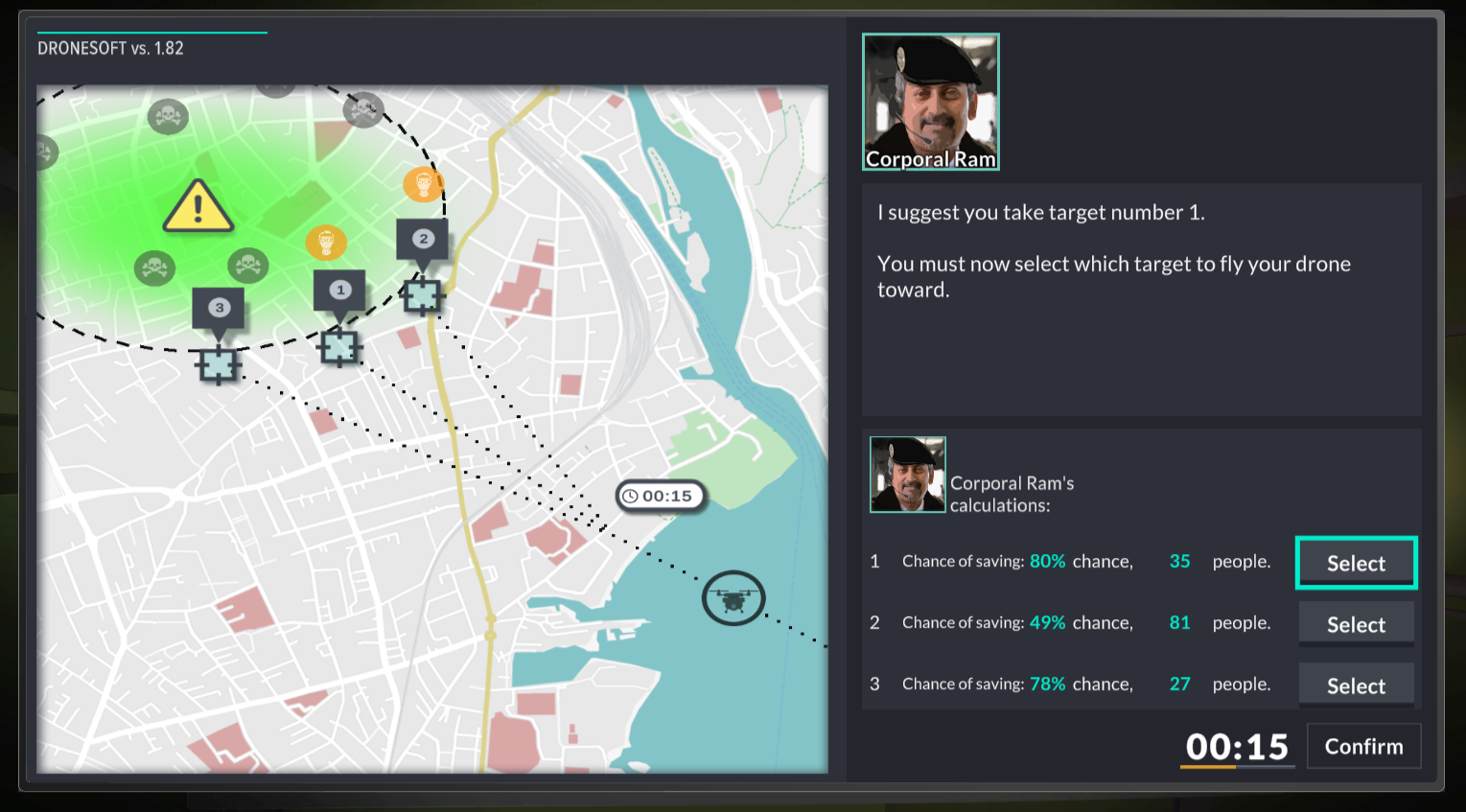}
\end{subfigure}%
\begin{subfigure}{.5\textwidth}
  \centering
  \includegraphics[width=.9\linewidth]{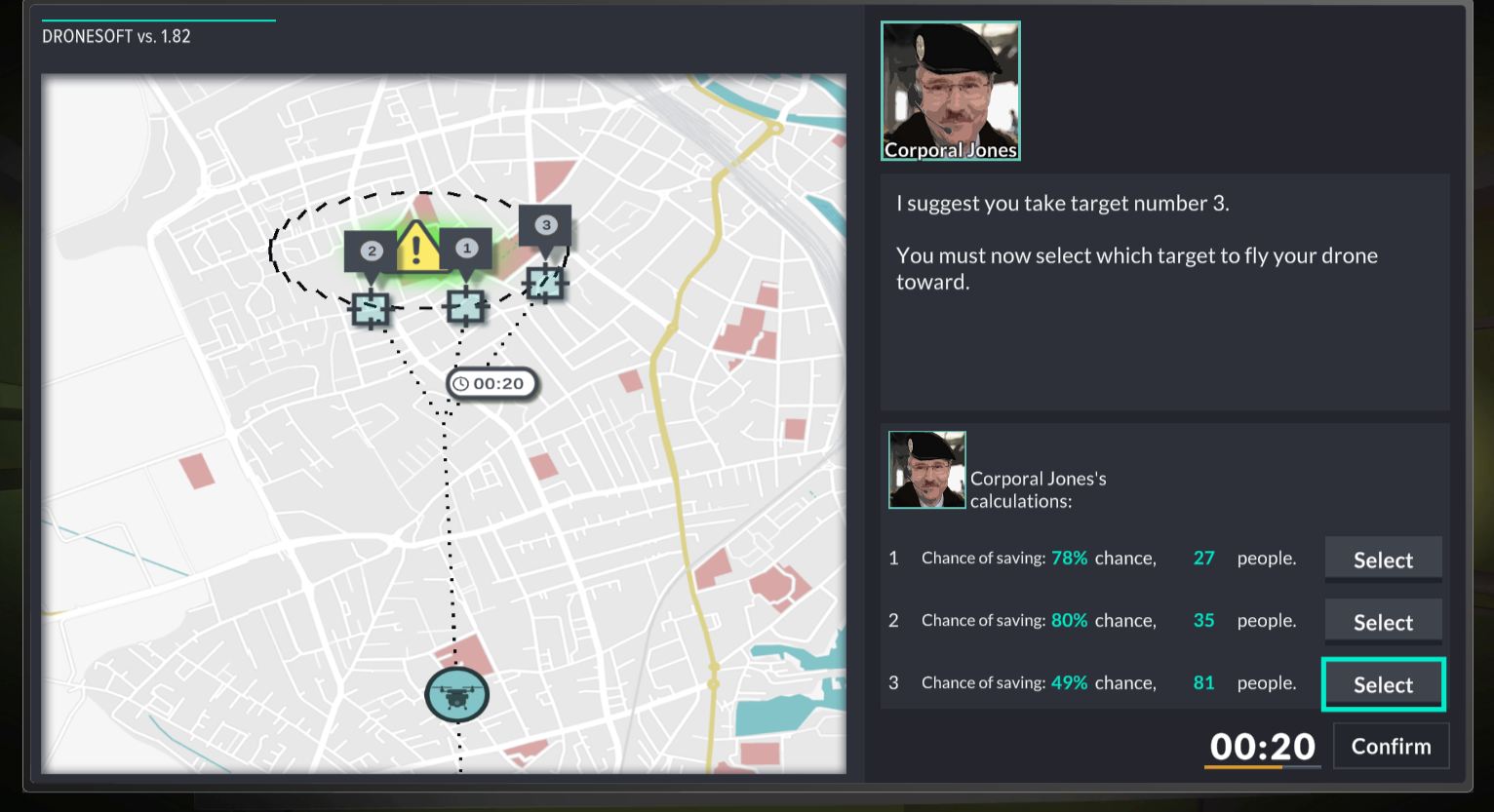}
\end{subfigure}
\caption{The screenshots show some design differences between the Indian (left) and the OECD (right) simulations for a human-in-the-loop setting and save scenario.}
\label{fig:test}
\end{figure}

\subsection{Participants}
\label{sec:participant}
Table \ref{tab:sample} presents an overview of the main characteristics of the two samples. The Table reveals that OECD participants are younger with larger share of females. OECD participants achieved lower education levels. Many more Indians studied math. 
Indian sample scored higher on affinity with technology. Indian participants are somewhat more risk-averse and more utilitarian. 

\begin{table}[ht]
    \centering
    \resizebox{.4\textwidth}{!}{
    \begin{tabular}{lccc}
    \toprule
    \textbf{Variable} & \textbf{OECD} & \textbf{India} &  \\
    & (n=351) & (n=302) & \\
    \midrule
   \textit{Age} & 25.4 ***    &   35.0*** &\\
     &   (7.8)     &  (7.9)&\\
  \textit{Gender} &  &  &  \\
\% females & 57.5*** &  23.8*** & \\
    & & &\\
   \textit{Education} & & & \\
   School, vocational  & 30 &13 &\\
   High school & 138  & 5& \\
   University & 183& 204& \\
   &&&\\
   \textit{Math study} & & & \\
   \% yes  & 36.1*** & 81.1***\\
   &&&\\
   \textit{Affinity to technology} & & &\\           & 4.23*** &  4.74***&\\
 &(0.86)  & (0.60)&\\
\textit{Risk measure} & & &\\
&4.01***          &      4.23***&\\
&(0.50)          &  (0.57)& \\
\textit{Utilitarianism score} & & & \\
&3.92***   &             4.83***&\\
&(0.91)  &                      (0.95)&\\ 
    \bottomrule
    \end{tabular}
    }
    \caption{Sample characteristics. Mean values and standard deviation in brackets.\\ Significance: *** p<0.01, ** p<0.05, * p<0.1; Mann-Whitney test}
    \label{tab:sample}
\end{table}

\section{Results}
\label{sec:results}
We present results of the analysis relevant for the posed research question. We compare
the results on trust, perceived responsibility, and reliance on the experts. For each of the dependent variable, we start with two-way mixed ANOVA for sample (OECD vs India, between-group) and expert type (AI vs. Human, within-group) and report effect sizes followed by random effect regressions with or without participants' characteristics as control variables. Finally, Figures visualize our main findings.

\subsection{Trust}
For capacity trust, we observe no effect of sample ($F=0.8$, $p>0.05$, $\eta^2=0.00010$),
a significant effect of expert type ($F=12.5$, $p<0.05$, $\eta^2=0.004$) and a significant interaction effect ($F=89.4$, $p<0.05$, $\eta^2=0.027$).
Post-hoc analysis suggests that sample effect is significant for both AI ($F=20.9$, $p<0.05$, $\eta^2=0.031$) and human expert ($F=14.9$, $p<0.05$, $\eta^2=0.022$)

For moral trust, we observe significant effect of sample ($F=41.6$, $p<0.05$, $\eta^2=0.04$),
a significant effect of expert type ($F=71.3$, $p<0.05$, $\eta^2=0.039$) and insignificant interaction.
Post-hoc pairwise t-tests show strong significant effects of expert type ($t=-8.41$, $p<0.01$)
and sample ($p<0.01$).
  
Finally, for the total trust there is a small significant effect of expert type ($F=7.2$, $p<0.05$, $\eta^2=0.002$), no effect of the sample, and the significant effect of the interaction ($F= 67.9$, $p<0.05$, $\eta^2=0.021$). Post-hoc analysis shows that the effect of the sample is significant for both AI ($F=4.97$, $p<0.05$, $\eta^2=0.008$) and human expert ($F=31.7$, $p<0.01$, $\eta^2=0.047$).

To analyze whether the sample characteristics might be responsible for the observed differences in trust, we ran the following linear random effect model:

\begin{equation}
Trust_i = \beta_0 +\beta_1\cdot Expert_i+\beta_2\cdot Sample_i+ \beta3\cdot Sample*Expert_i+a_i+e_i
\end{equation}
\label{eq:regress_trust}

$Trust$ is the respective trust score (the average score of the respective items for capacity, moral, or overall trust), $Expert$ is a dummy variable for Human vs. AI expert, $Sample$ is a dummy variable for Indian vs. OECD sample, $Expert*Sample$ is the interaction term, $a_i$ is the individual random effect, and $e_i$ is the error term.
We ran the specification \ref{eq:regress_trust} with and without controls. The control variables include all the sample characteristics specified in Table \ref{tab:sample}.

\begin{table}
\begin{center}
\begin{tabular}{l c c c c c c }
\hline
 & \multicolumn{2}{c}{Capacity trust} & \multicolumn{2}{c}{Moral trust} & \multicolumn{2}{c}{Total trust} \\
 
\hline
(Intercept)               & $5.22^{***}$  & $3.08^{***}$  & $4.99^{***}$  & $3.31^{***}$  & $5.11^{***}$  & $2.87^{***}$  \\
                          & $(0.07)$      & $(0.47)$      & $(0.09)$      & $(0.59)$      & $(0.07)$      & $(0.46)$      \\
expert:human           & $0.24^{***}$  & $0.24^{***}$  & $0.65^{***}$  & $0.65^{***}$  & $0.44^{***}$  & $0.44^{***}$  \\
                          & $(0.06)$      & $(0.06)$      & $(0.11)$      & $(0.11)$      & $(0.06)$      & $(0.06)$      \\
sample:OECD                & $0.44^{***}$  & $0.62^{***}$  & $-0.64^{***}$ & $-0.59^{***}$ & $0.21^{*}$    & $0.42^{***}$  \\
                          & $(0.09)$      & $(0.13)$      & $(0.13)$      & $(0.17)$      & $(0.09)$      & $(0.12)$      \\
human*OECD & $-0.76^{***}$ & $-0.76^{***}$ & $-0.02$       & $-0.02$       & $-0.67^{***}$ & $-0.67^{***}$ \\
                          & $(0.08)$      & $(0.08)$      & $(0.15)$      & $(0.15)$      & $(0.08)$      & $(0.08)$      \\
age                       &               & $-0.00$       &               & $-0.01^{*}$   &               & $-0.00$       \\
                          &               & $(0.01)$      &               & $(0.01)$      &               & $(0.00)$      \\
gender:Male                &               & $-0.25^{**}$  &               & $-0.05$       &               & $-0.20^{*}$   \\
                          &               & $(0.09)$      &               & $(0.11)$      &               & $(0.09)$      \\
gender:Other               &               & $-0.01$       &               & $-0.08$       &               & $-0.16$       \\
                          &               & $(0.45)$      &               & $(0.56)$      &               & $(0.44)$      \\
education:School        &               & $0.04$        &               & $-0.14$       &               & $0.10$        \\
                          &               & $(0.17)$      &               & $(0.22)$      &               & $(0.17)$      \\
education:University    &               & $0.06$        &               & $-0.25$       &               & $0.08$        \\
                          &               & $(0.11)$      &               & $(0.14)$      &               & $(0.11)$      \\
math:Yes                &               & $-0.13$       &               & $-0.07$       &               & $-0.13$       \\
                          &               & $(0.09)$      &               & $(0.11)$      &               & $(0.09)$      \\
affinity\_score                  &               & $0.26^{***}$  &               & $0.32^{***}$  &               & $0.27^{***}$  \\
                          &               & $(0.06)$      &               & $(0.07)$      &               & $(0.06)$      \\
risk\_score                 &               & $0.07$        &               & $0.03$        &               & $0.07$        \\
                          &               & $(0.07)$      &               & $(0.09)$      &               & $(0.07)$      \\
utilitarian\_score          &               & $0.20^{***}$  &               & $0.19^{***}$  &               & $0.20^{***}$  \\
                          &               & $(0.04)$      &               & $(0.05)$      &               & $(0.04)$      \\
\hline
R$^2$                     & 0.08          & 0.12          & 0.08          & 0.12          & 0.05          & 0.10          \\
Adj. R$^2$                & 0.07          & 0.11          & 0.08          & 0.11          & 0.05          & 0.09          \\
Num. of subjects               & 649          & 649          & 649          & 649          & 649          & 649          \\
\hline
\multicolumn{7}{l}{\scriptsize{Random effect model. $^{***}p<0.001$, $^{**}p<0.01$, $^*p<0.05$}}
\end{tabular}
\caption{Regression output: Trust}
\label{tab:regress_trust}
\end{center}
\end{table}

In India, human experts score higher than AI on all trust scales, as captured by the positive and significant coefficient $expert:human$. In OECD, participants trust human less than AI in terms of capacity trust, as captured by the negative sum of coefficients $expert:human$ and $human*OECD$, but more than AI in terms of moral trust which is captured by the positive sum of coefficients $expert:human$ and $human*OECD$.

Among the control variables, $gender:Male$ is significant for capacity and overall trust : males seem to trust the experts less than females. Besides gender, affinity to technology and utilitarianism have an impact on trust. These variables, however, do not seem to alter the differences in trust between the OECD and India samples. 

\begin{figure}
    \centering
    \includegraphics[width=0.7\textwidth]{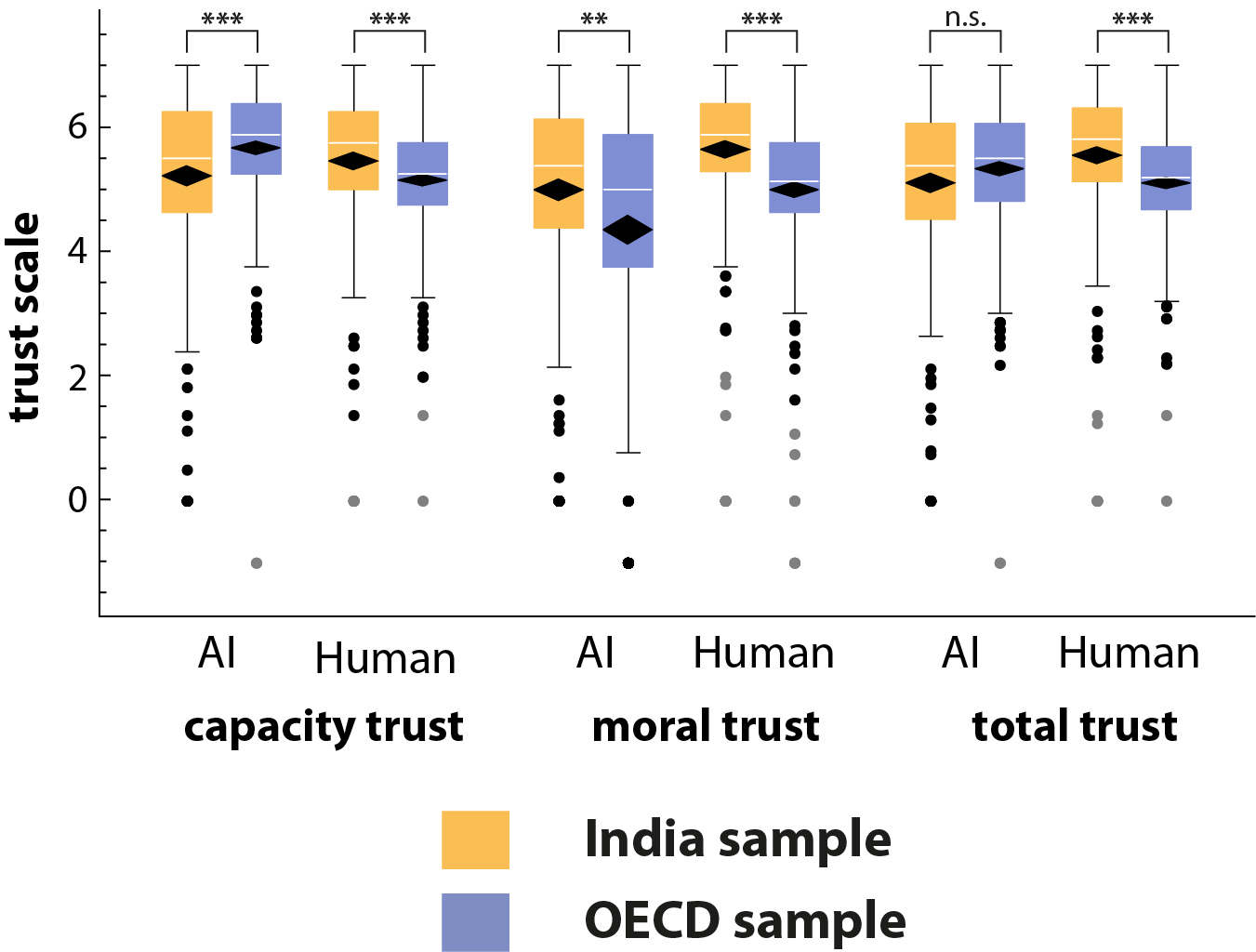}
    \caption{Box plots of trust scales: mean values (diamonds) and median values (white bar). Significance: *** p<0.01, ** p<0.05, * p<0.1; t-test} 
    \label{fig:trust1}
\end{figure}

Figure \ref{fig:trust1} summarizes the comparison of the trust scales. The Figure reveals several important observations. In India, AI scored lower in capacity trust than in OECD. Human expert scored higher in capacity trust in India than in OECD. Moral trust for both expert types is lower in Indian than in OECD. Indians trusted human experts overall more often than in OECD, whereas AI scored similarly on overall trust in both populations.

\begin{figure}
    \centering
    \includegraphics[width=0.7\textwidth]{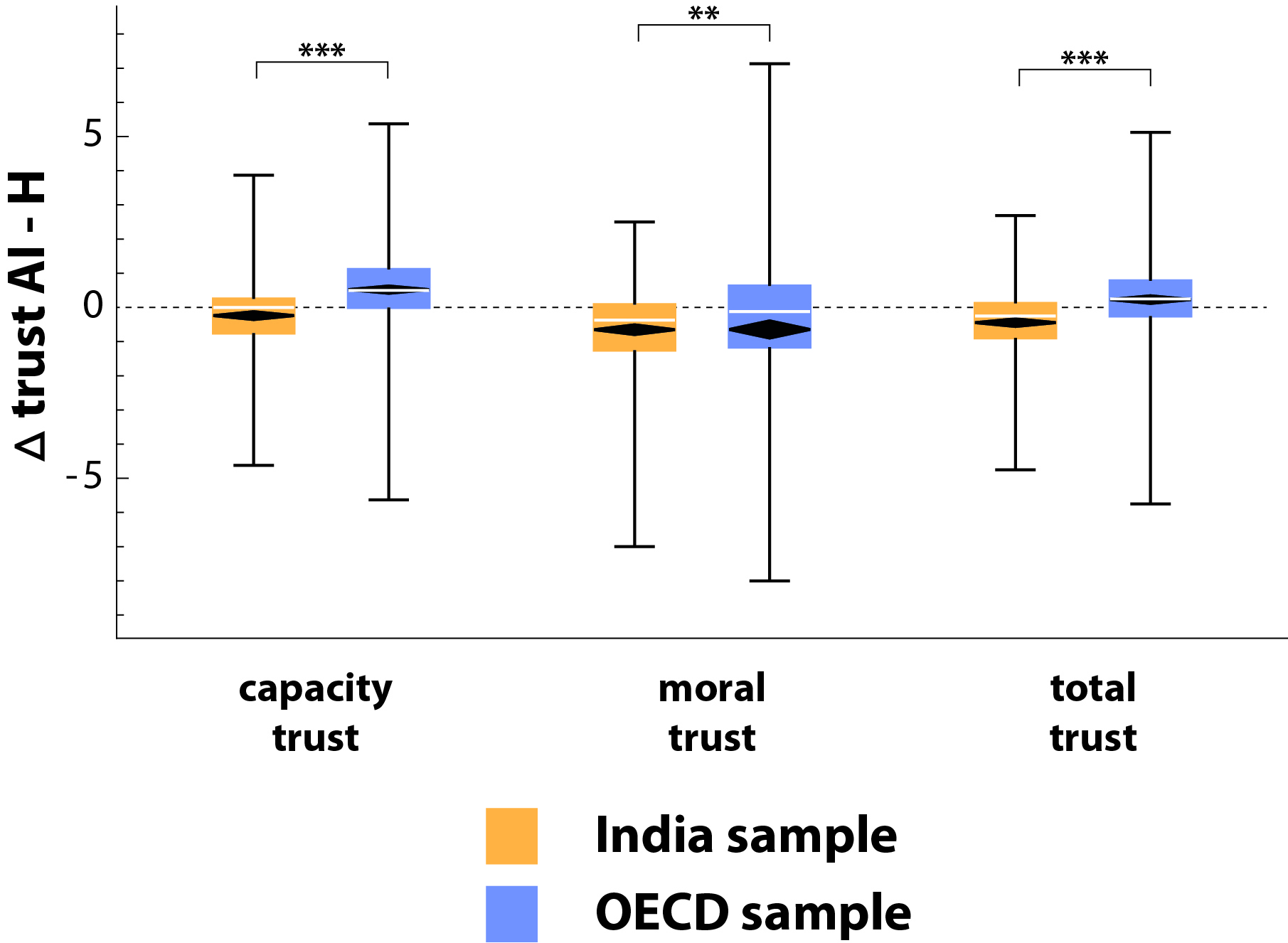}
    \caption{Box plots of trust scales: mean values (diamonds) and median values (white bar). Significance: *** p<0.01, ** p<0.05,* p<0.1; t-test} 
    \label{fig:trust2}
\end{figure}

 Let's now turn to difference in perceived trust between AI and human expert across Indian and OECD samples. Figure \ref{fig:trust2} depicts the difference in the average score across items for respective subscale of trust. The Figure shows that Indians perceive both expert types to be equally capable in solving the task. OECD participants believe AI to be more capable than a human operator; the result consistent with our previous findings in \cite{Tolmeijer:2022}. Both populations consider humans to be more morally trustworthy than AI (diamonds below zero dashed line), the difference in moral trust between expert types is larger for OECD sample. Finally, Indians assign slightly more overall trust to humans (negative difference, below zero dashed line) than to AI, whereas for OECD participants the opposite effects of capacity and moral trust cancel each other in the overall trust measure.
 
\subsection{Responsibility}

We now turn to perceived responsibility: For self-responsibility, we observe no significant effects from two-way ANOVA. For responsibility assigned to experts, we observe significant effects of sample ($F=137.7$, $p<0.001$,           $\eta^2=0.136$),  expert type ($F=295$, $<0.001$, $\eta^2=0.104$) and the interaction ($F=65$, $p<0.001$, $\eta^2=0.025$). Finally, for responsibility assigned to sellers and programmers, one way ANOVa reveals significant effects of the sample ($F=128.5$, $p<0.001$, $\eta^2=0.09$ and $F=185.3$, $p<0.001$, $\eta^2=0.125$)

To analyze whether the sample characteristics might be responsible for the observed differences in responsibility, we ran the following linear random effect model:

\begin{equation}
Responsibity_i = \beta_0 +\beta_1\cdot Expert_i+\beta_2\cdot Sample_i+ \beta3\cdot Sample*Expert_i+a_i+e_i
\end{equation}
\label{eq:regress_responsibility}

$Responsibility$ is the respective responsibility measure (own responsibility and the responsibility assigned to the expert), $Expert$ is a dummy variable for Human vs. AI expert, $Sample$ is a dummy variable for Indian vs. OECD sample, $Expert*Sample$ is the interaction term, $a_i$ is the individual random effect, and $e_i$ is the error term.
We ran the specification \ref{eq:regress_responsibility} with and without controls. The control variables include all the sample characteristics specified in Table \ref{tab:sample}.

Table \ref{tab:reg_resp} presents the regression output. The Table shows that there are no differences in the responsibility participants assign to themselves across samples or expert types.
As regards the AI experts, Indian participants hold them much more responsible that OECD participants (negative significant coefficient $sample: OECD$).
Indian participants assign higher responsibility to programmers and sellers of AI. Gender, affinity to technology, and utilitarianism have similar effects as for the trust scales. Risk score and education seem to play a role, however, not in all specifications.

\begin{table}
\begin{center}
\begin{tabular}{l c c c c c c c c }
\hline
 & \multicolumn{2}{c}{Responsibility: self}  & \multicolumn{2}{c}{Responsibility: expert} & \multicolumn{2}{c}{*Responsibility: programmer} & \multicolumn{2}{c}{*Responsibility: seller} \\
\hline
(Intercept)               & $5.85^{***}$ & $3.41^{***}$ & $4.49^{***}$  & $1.31$        & $4.87^{***}$  & $2.99^{***}$  & $4.72^{***}$  & $2.39^{***}$  \\
                          & $(0.07)$     & $(0.57)$     & $(0.10)$      & $(0.67)$      & $(0.07)$      & $(0.57)$      & $(0.07)$      & $(0.58)$      \\
expert:human            & $0.08$       & $0.08$       & $0.61^{***}$  & $0.61^{***}$  &               &               &               &               \\
                          & $(0.05)$     & $(0.05)$     & $(0.10)$      & $(0.10)$      &               &               &               &               \\
sample:OECD                & $0.10$       & $0.25$       & $-1.89^{***}$ & $-1.67^{***}$ & $-1.31^{***}$ & $-1.34^{***}$ & $-1.10^{***}$ & $-0.86^{***}$ \\
                          & $(0.10)$     & $(0.15)$     & $(0.13)$      & $(0.18)$      & $(0.10)$      & $(0.14)$      & $(0.10)$      & $(0.15)$      \\
human*OECD & $-0.12$      & $-0.12$      & $1.09^{***}$  & $1.09^{***}$  &               &               &               &               \\
                          & $(0.06)$     & $(0.06)$     & $(0.13)$      & $(0.13)$      &               &   $(0.13)$      &                    &               \\
age                       &              & $0.00$       &               & $-0.01$       &               & $-0.00$       &               & $-0.00$       \\
                          &              & $(0.01)$     &               & $(0.01)$      &               & $(0.01)$      &               & $(0.01)$      \\
gender:Male                &              & $-0.31^{**}$ &               & $-0.23$       &               & $-0.31^{**}$  &               & $-0.12$       \\
                          &              & $(0.11)$     &               & $(0.13)$      &               & $(0.11)$      &               & $(0.11)$      \\
gender:Other               &              & $-1.04$      &               & $-2.04^{**}$  &               & $0.78$        &               & $0.12$        \\
                          &              & $(0.54)$     &               & $(0.64)$      &               & $(0.55)$      &               & $(0.55)$      \\
education:        &              & $0.19$       &               & $-0.03$       &               & $-0.60^{**}$  &               & $-0.65^{**}$  \\
            School              &              & $(0.21)$     &               & $(0.25)$     &               & $(0.21)$      &               & $(0.21)$      \\
education:    &              & $0.13$       &               & $-0.31^{*}$   &               & $-0.64^{***}$ &               & $-0.19$       \\
        University                  &              & $(0.14)$     &               & $(0.16)$      &               & $(0.14)$      &               & $(0.14)$      \\
math:Yes                   &              & $-0.04$      &               & $0.05$        &               & $0.10$        &               & $-0.08$       \\
                          &              & $(0.11)$     &               & $(0.13)$      &               & $(0.11)$      &               & $(0.11)$      \\
affinity\_score                  &              & $0.34^{***}$ &               & $0.21^{**}$   &               & $0.15^{*}$    &               & $0.07$        \\
                          &              & $(0.07)$     &               & $(0.08)$      &               & $(0.07)$      &               & $(0.07)$      \\
risk\_score                &              & $0.23^{**}$  &               & $0.34^{**}$   &               & $0.23^{*}$    &               & $0.16$        \\
                          &              & $(0.09)$     &               & $(0.10)$      &               & $(0.09)$      &               & $(0.09)$      \\
utilitarian\_score          &              & $-0.01$      &               & $0.29^{***}$  &                & $0.22^{***}$  &               & $0.36^{***}$  \\
                          &              & $(0.05)$     &               & $(0.06)$      &               & $(0.05)$      &               & $(0.05)$      \\
\hline
R$^2$                     & 0.00         & 0.04         & 0.29          & 0.32          & 0.13          & 0.18          & 0.09          & 0.14          \\
Adj. R$^2$                & 0.00         & 0.03         & 0.28          & 0.32          & 0.12          & 0.17          & 0.09          & 0.13          \\
Num. of subjects              & 649         & 649         & 649           & 649           & 649           & 649          & 649           & 649          \\
\hline
\multicolumn{9}{l}{Random effects model; *OLS model. \scriptsize{$^{***}p<0.001$, $^{**}p<0.01$, $^*p<0.05$}}
\end{tabular}
\caption{Regression output: Responsibility}
\label{tab:reg_resp}
\end{center}
\end{table}

\begin{figure}[h]
    \centering
    \includegraphics[width=0.8\textwidth]{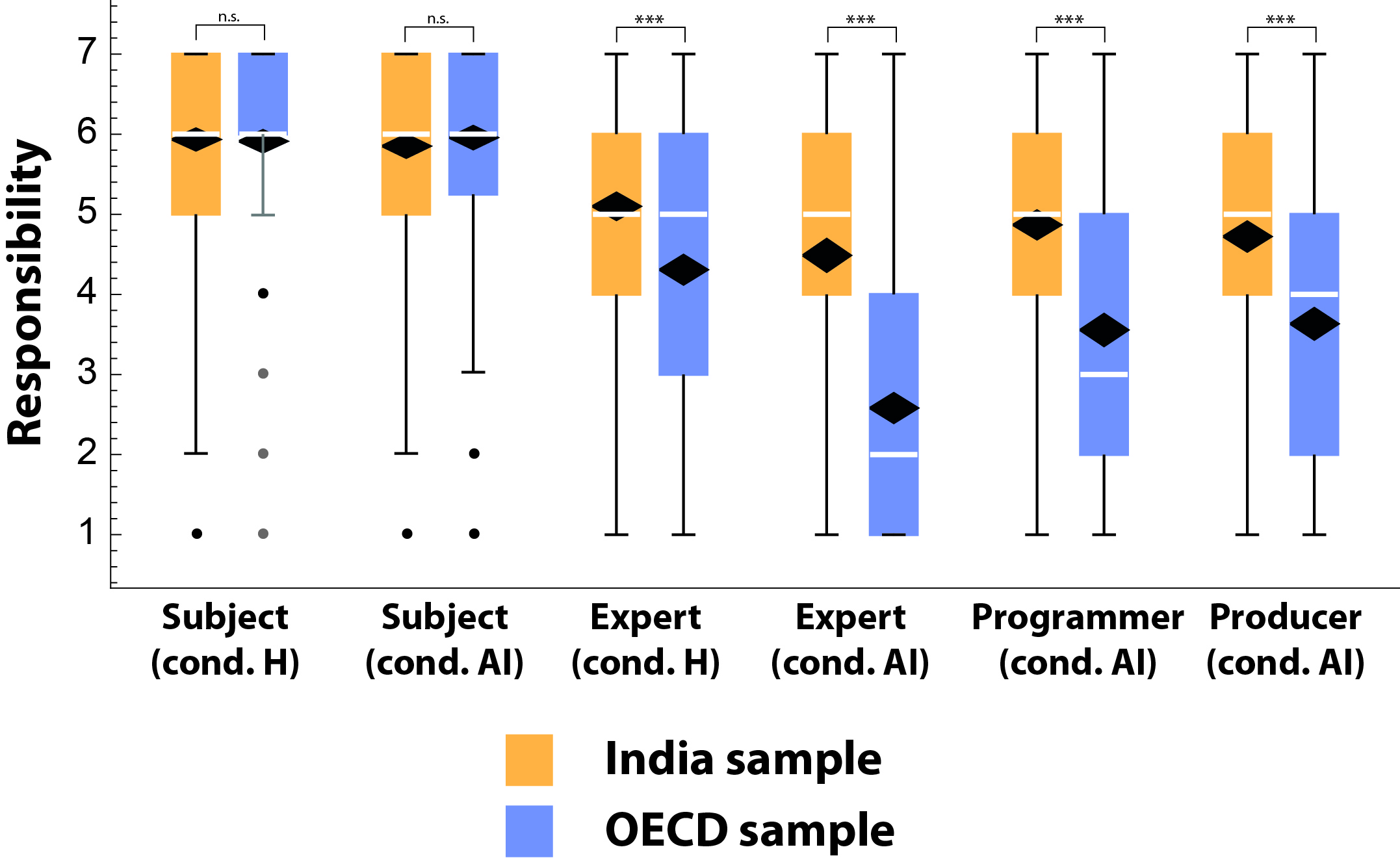}
    \caption{Boxplots of the assigned responsibility scores.  A responsibility score of 1 indicates the participant thought the entity to be `not responsible at all', while 7 implies they found them to be `very responsible'.}
    \label{fig:resp}
\end{figure}

Figure \ref{fig:resp}
presents the responsibility scores for decision makers themselves, experts by types, and for programmers and developers of the AI. The Figure shows that participants in both populations felt very responsible for the outcomes. Indians and OECD participants also similarly assessed the responsibility of the human expert. Interestingly, in contrast to OECD participants, Indians assigned higher responsibility scores for AI expert, the programmers and the developers. These findings suggest that Indians hold all parties behind the outcomes of the decisions responsible. In other words, for Indian participants both human and AI experts seem to have the same degree of agency for the task at hand. 

\subsection{Reliance}

\begin{figure}
\centering
\includegraphics[width=0.8\textwidth]{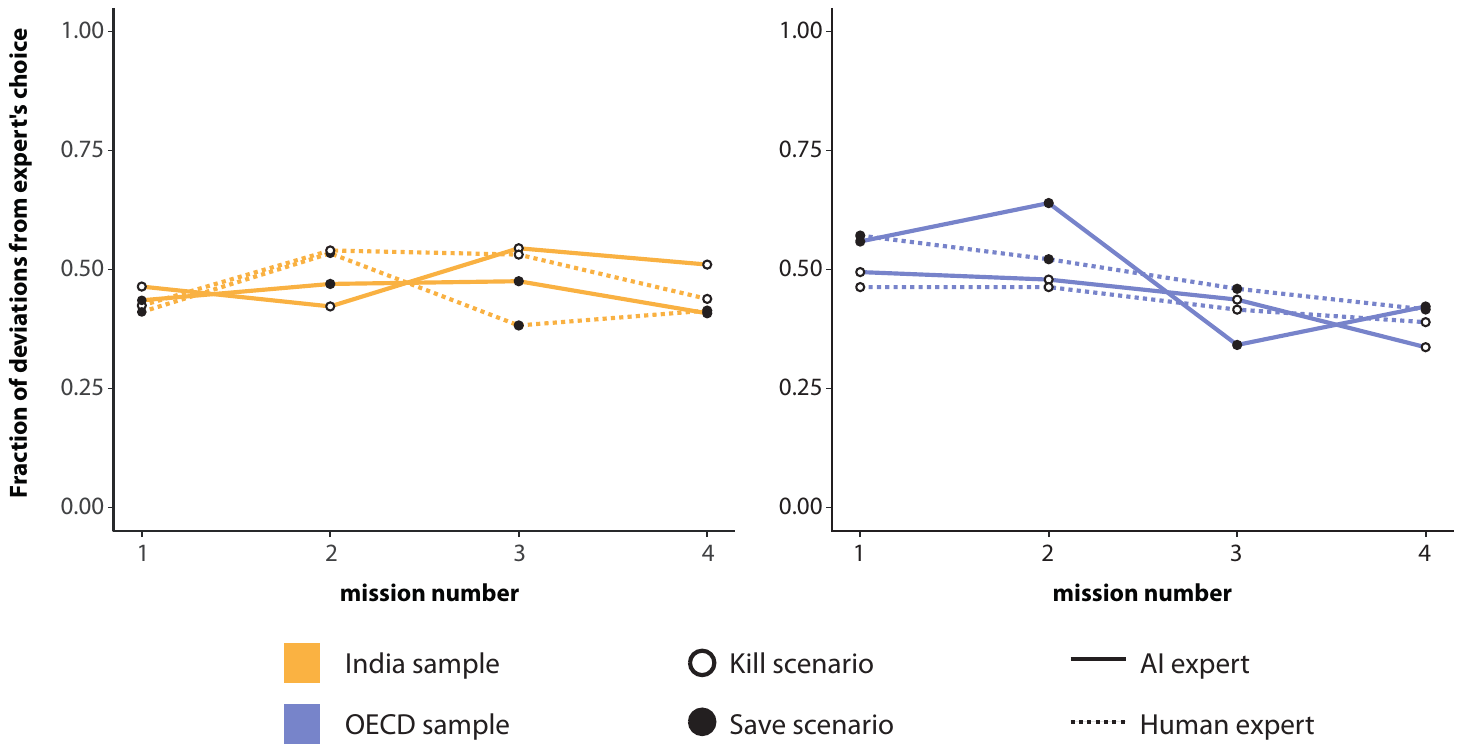}
\caption{Reliance on expert's advice. Higher values on Y-axis indicate lower rate of reliance.}
\label{fig:reliance}
\end{figure}

\begin{table}[!ht]
\begin{center}
\begin{tabular}{l c c }
\hline
 & OECD & India \\
\hline
(Intercept)         & $0.49^{***}$  & $0.46^{***}$ \\
                    & $(0.02)$      & $(0.03)$     \\
expert:human     & $-0.00$       & $-0.01$      \\
                    & $(0.02)$      & $(0.02)$     \\
scenario: ``save''   & $0.06^{**}$   & $-0.04$      \\
                    & $(0.02)$      & $(0.02)$     \\
mission2 & $0.00$        & $0.06^{*}$   \\
                    & $(0.03)$      & $(0.03)$     \\
mission3 & $-0.10^{***}$ & $0.05$       \\
                    & $(0.03)$      & $(0.03)$     \\
mission4 & $-0.13^{***}$ & $0.01$       \\
                    & $(0.03)$      & $(0.03)$     \\
\hline
R$^2$               & 0.02          & 0.00         \\
Adj. R$^2$          & 0.02          & 0.00         \\
Num. obs.           & 2808          & 2307         \\
R$^2$                            & 0.02          & 0.00         \\
Adj. R$^2$                       & 0.02          & 0.00         \\
Num. of subjects                       & 351          & 302         \\
\hline
\multicolumn{3}{l}{Random effect model. \scriptsize{$^{***}p<0.001$, $^{**}p<0.01$, $^*p<0.05$}}
\end{tabular}
\caption{Reliance on expert's advice}
\label{tab:reg_reliance}
\end{center}
\end{table}

We conclude the results section by looking at reliance: how often the participants followed the advice from the expert. Figure \ref{fig:reliance} compared reliance on expert's advice by and expert type and scenario ("kill" vs. "save") between OECD and Indian sample. As the Figure shows, OECD and Indian sample differ in the rate of reliance on experts in kill and save scenarios.
OECD participants follow expert's advice more often in ``save'' than in ``kill''. Indian participants seem to do the opposite:
they follow experts' advice more often in ``kill'' than in ``save''.
Overall there are no tangible differences between reactions toward human and AI's advice.

We conclude this section by regression rate of deviations from expert's advice on expert type and scenario:

\begin{equation}
Deviation_i = \beta_0 +\beta_1\cdot Expert_i+\beta_2\cdot Scenario+\beta_3\cdot Mission+a_i+e_i
\end{equation}
\label{eq:regress_reliance}
Where \textit{Deviation} is a binary variable with takes a value 1 of a participant does not follow the advice from the expert and 0 otherwise;
$Expert_i$ is a dummy for Human and AI expert; $Scenario$ is a dummy for ``kill'' and ``save'' condition,
$Mission$ is a fixed effect of a mission; $a_i$ is the individual random effect; $e_i$ is the error term.

Table \ref{tab:reg_reliance} presents the estimation results for OECD and Indian samples.

As Table \ref{tab:reg_reliance} shows, there is no difference in the rate of reliance on AI and human advice both in OECD and India.
(Insignificant coefficient $expert:human$). OECD participants follow advice slightly less often in ``save'' (+6\% more deviations from the provided advice). Scenario type does not seem to play a role for Indian participants (negative but insignificant coefficient $scenario:save$).
The level of reliance goes up in the later stages of the game (mission 3 and 4) for OECD participants and remains relatively stable for Indian throughout the game. 

\section{Discussion and conclusion}
\label{sec:discussion}

We find small, though significant differences in overall trust and moral trust across populations. The difference in capacity trust seems the most pronounced (Figure \ref{fig:trust2}), with participants from India vesting more trust in humans, whereas participants from OECD countries more in AI advisors. The responsibility assumed by participants was high in both conditions and did not differ significantly across samples (Figure \ref{fig:resp}). However, whereas OECD participants were relatively unwilling to attribute responsibility to an AI advisor, its programmer, or producer, (none significantly above the midpoint of the scale), mean responsibility attributions for all three was high in India  (and significantly above the midpoint). For reliance on expert advice, we found an interaction between scenario type and culture , though little in the vein of a profound preference of advice for either type of expert (human or AI) in either culture.   
Overall then, there is considerable convergence across cultures, with the exception of the disposition of Indians to hold AI programmers and producers as well as the AI itself responsible to much higher degrees than OECD participants do. What could explain this finding? The pronounced difference regarding other humans involved responsible seems to be consistent with Hofstede’s classic West/East observation: Whereas in OECD countries, participants – being the last humans involved in the action sequence – shoulder the responsibility predominantly themselves. In India, with its anti-individualistic and collectivist tendencies, responsibility for the consequences is distributed across all humans whose actions (at an earlier or later stage) influenced the resulting outcome. 
What remains are the divergent findings concerning responsibility attribution to the AI. First, it must be noted that, as regards the West, our results are somewhat out of tune with some extant research which also explored attributions of moral responsibility and blame to AI-driven agents. \cite{kneer2021playing} and \cite{stuart2021}, whose studies also explore responsibility in contexts of risk and uncertainty report a rather pronounced disposition for US samples to hold AI-agents responsible, whereas the Westerners in our sample did not. One explanation for this disparity is that, in our experiment, the human participants had the “last word” on the decision, and the AI experts merely provided advice which could be overridden. This finding might put some of the worries of moral philosophers about “responsibility gaps” and “retribution gaps” to rest (see e.g. \citep{matthias2004, sparrow2007, danaher2016}; for a review see \cite{santoni2018meaningful}).
For OECD participants, it seems clear that they carry moral responsibility in human-AI collaborations, and that – in virtue of collaborating with AI – the very possibility of attributing responsibility does not simply vanish into thin air. This leaves us with the findings of Indian subjects who attribute a high degree of responsibility to the AI.  One hypothesis could be that the collective (or the confines of the group agent) that is deemed responsible includes artificial agents, too. Naturally, to date, and in particular given the dire need of more research with Indian participants, it is difficult to assess how plausible this hypothesis is. Responsibility attribution in human-AI teams across the East/West divide, do, however, constitute an interesting avenue for further research.

\section*{Acknowledgments} 
This work is partially funded by Armasuisse Science and Technology (S+T), via the Swiss Center for Drones and Robotics of the Department of Defense, Civil Protection and Sport (DDPS). This work is partially funded by National Research Program (NRP) 77 `Digital Transformation' of the Swiss National Science Foundation under project number 407740\_187494.
We thank Koboldgames for their support during the design process of the simulation. 

\clearpage

\bibliographystyle{abbrvnat}
\bibliography{references}

\end{document}